\documentclass[11pt]{article}
\usepackage{amssymb,latexsym}
\newtheorem{theo}{Theorem}
\newtheorem{definition}[theo]{Definition}
\def\a{\alpha}
\def\b{\beta}
\def\g{\gamma}
\def\d{\delta}
\def\t{\theta}
\def\P{{\bf P}}
\def\R{{\bf R}}
\def\p{{\bf p}}
\def\r{{\bf r}}

\def\l{{\ldots}}
\def\n{{\noindent}}
\def\[{[\![}
\def\]{]\!]}
\def\Z{{\bf Z}}
\def\N{{\bf N}}
\def\s{{\smallskip }}
\def\n{{\noindent }}
\def\nn{{\nonumber }}

\begin{document}
\begin{center} 
{\Large \bf Wigner quantum systems (Lie superalgebraic approach)}\\[5mm]
T. D. Palev \\Institute for Nuclear
Research and Nuclear Energy, 1784 Sofia, Bulgaria
\\  tpalev@inrne.acad.bg \\[2mm]
N. I. Stoilova
\\ Mathematical Physics Group, Department of Physics,
Technical University of \\Clausthal, Leibnizstrasse 10, D-38678
Clausthal-Zellerfeld, Germany\\Institute for Nuclear
Research and Nuclear Energy, 1784 Sofia, Bulgaria
\\ ptns@pt.tu-clausthal.de, stoilova@inrne.bas.bg 
\end{center}

\begin{abstract}
      We present three groups of examples of Wigner Quantum
      Systems related to the Lie superalgebras $osp(1/6n)$,
      $sl(1/3n)$ and $sl(n/3)$ and discuss shortly their physical
      features. In the case of $sl(1/3n)$ we indicate that the
      underlying geometry is noncommutative.

\end{abstract}

\section{Introduction}

We present some examples of noncanonical
nonrelativistic quantum systems, referred to as Wigner quantum
systems. The corresponding to them statistics are the so called $A$-
and  $B$-statistics.

The justification for studying such more general quantum systems
is based on two key
issues.  The first one stems from the ideas of Wigner
in his 1950's work {\it Do the equations of motion determine the
quantum mechanical commutation relations?}~\cite{Wigner}. In this paper he
has shown that the canonical quantum statistics may, in principle, be
generalized in a logically consistent way.  Wigner demonstrated this on an
example of a one-dimensional oscillator with a Hamiltonian
($m=\omega=\hbar=1$) $H={1\over 2}(p^2 + q^2)$.  Abandoning the canonical
commutation relations (CCRs) $[p,q]=-i,$ he was searching for all
operators $q$ and $p$, such that the "classical" Hamiltonian equations
${\dot p}=-q,~~~ {\dot q}=p$ were identical with the Heisenberg equations
${\dot p}=-i[p,H],~{\dot q}=-i[q,H].$ Apart from the canonical solution he
found infinitely many other solutions. In this relation Wigner noted that
the Hamiltonian Eqs.  and the Heisenberg Eqs. have a more direct physical
significance than the CCRs. Therefore, concluded Wigner, it is logically
justified to postulate them from the very beginning, instead of the CCRs.
On the ground of these considerations, {\it a Wigner quantum system} (WQS)
is defined as a quantum system for which the Hamiltonian Eqs.  are
identical to the Heisenberg Eqs.. Then, in general, the CCRs do not hold.

An important further step was the observation that all solutions
found by Wigner turned to be different representations of one
pair of para-Bose (pB) creation and annihilation operators
(CAOs)~\cite{Ohnuki}.
Let us note that in 1950 the parastatistics was still
unknown. It was introduced three years later by Green~\cite{Green} as a
possible generalization of Bose and Fermi statistics in quantum field
theory.

The second key issue is based on the observation that any $n$
pairs of Bose operators $b_1^\pm,\l,b_n^\pm$ generate a
representation, the Bose representation $\rho_b$, of the
orthosymplectic Lie superalgebra (LS) $osp(1/2n)=B(0/n)$.

The Cartan-Weyl generators (in this representation) are the Bose
operators and all of their anticommutators
$\{b_i^\xi,b_j^\eta\},\;\xi,\eta=\pm.$ Denote by
$B_1^\pm,\l,B_n^\pm$ those generators of $B(0/n)$, which in the
Bose representation coincide with the Bose operators,
$\rho_b(B_i^\pm)=b_i^\pm$. Then the representation independent
operators $B_1^\pm,\l,B_n^\pm$ generate $B(0/n)$
~\cite{Ganchev}   and satisfy the triple relations :
\begin{equation}
\label{pB}
[\{B_i^\xi,B_j^\eta\},B_k^\epsilon]=(\epsilon-\xi)\delta_{ik}B_j^\eta
+(\epsilon-\eta)\delta_{jk}B_i^\xi.  \end{equation} The
operators~(\ref{pB}) are known in quantum field theory. These are the
para-Bose operators, mentioned above, which generalize the statistics of
the tensor fields~\cite{Green}. This observation is important. It casts a
bridge between two very different algebraical structures: para-Bose
statistics and the representation theory of Lie superalgebras.
 The more precise statement is
that the representation theory of $n$ pairs of para-Bose CAOs is
completely equivalent to the representation theory of the
orthosymplectic Lie superalgebra $B(0/n)$~\cite{Ganchev}. A
similar statement holds also for Fermi statistics and its
generalization, the para-Fermi statistics~\cite{Green}: any $n$
pairs of para-Fermi (pF) operators $F_1^\pm,\l,F_n^\pm$ generate
the orthogonal Lie algebra $B_n$~\cite{Kamefuchi}. Note that both
$B_n$ and $B(0/n)$ belong to the class {\it B} of the basic Lie
superalgebras ~\cite{Kac}. Therefore the pF and the pB statistics
(and in particular the Fermi and the Bose statistics) can be
called $B-$statistics.

On the ground of this observation statistics related to the other
infinite classes of basic Lie superalgebras, namely $A-$, $C-$
and $D-$statistics~\cite{Palev1}-\cite{Palev4} were introduced.
So far only the $A-$statistics were studied in more
detail~\cite{Palev5},\cite{Palev6}. By definition the statistics
of a quantum system (which has a classical analogue) is said to
be $A-$ (resp. $B-$, $C-$, $D-$) statistics if

\n $\bullet~$ The system is a Wigner quantum system;

\noindent $\bullet~$ The position and the momentum operators
$\r_\a$ and $\p_\a, \; \a=1,\ldots,n,$ of the system generate (a
representation of) a Lie superalgebra from the class $A$ (resp.
$B$, $C$, $D$).

Throughout the paper we use the notation: $\Z$ - the set of all
integers; $\Z_2=\{\bar{0},\bar{1}\}$ - the ring of all integers
modulo 2; $\N$ - all positive integers. Furthermore:
\begin{eqnarray}
&&[p;q]=\{p,p+1,p+2,\l,q-1,q\},\quad \hbox{for}\quad  p\le q\in \Z ;\\
&&\t_i=\cases {{\bar 0}, & if $\; i=0,1,2, \ldots , n$,\cr
               {\bar 1}, & if $\; i=n+1,n+2,\ldots ,n+m$,\cr }; \quad
\t_{ij}=\t_i+\t_j; \\ &&[a,b]=ab-ba,\;\; \{a,b\}=ab+ba,
\;\;\[a,b\]=ab-(-1)^{\deg(a)\deg(b)}ba, \end{eqnarray} where
$deg(a)\in\Z_2$ is the degree of the homogeneous element $a$ from the
superalgebra.

\section{Wigner quantum systems. Classes of solutions}

\begin{definition}
A system with a Hamiltonian
$ H_{tot}=\sum_{k=1}^N {\p_k^2\over 2m_k }+
V(\r_1,\r_2,\ldots,\r_N), $ is said to be a Wigner quantum system
\cite{Palev7} if the following conditions hold:

{\bf C1.} The state space $W$ is a Hilbert space. The observables
are Hermitian (selfadjoint) $~~~~~~~~~~~$ operators in $W$.

{\bf C2.} The Hamiltonian equations coincide with the Heisenberg
equations (as operator $~~~~~~~~~~~$ equations in $W$).

{\bf C3.} The description is covariant with respect to the
Galilean group.
\end{definition}

Here we consider a system of $n+1$ particles with mass $m$ and a
Hamiltonian:

\begin{equation}\label{Ham}
H_{tot}=\sum_{A=1}^{n+1}{(\P_A)^2 \over 2m}
+{m\omega^2\over{2(n+1)}}\sum_{A<B=1}^{n+1}
(\R_A-\R_B)^2.
\end{equation}
We proceed to show how this system can be turned into a
noncanonical WQS in three different ways, related to the Lie
superalgebras $osp(1/6n)$, $sl(1/3n)$ and $sl(n/3)$.

Introduce centre of mass (CM) and internal
variables~\cite{Moshinsky} ($\a=1,\ldots,n$):
\begin{equation}
\R={\sum_{A=1}^{n+1}\R_A \over{n+1}}, \;
   \P=\sum_{A=1}^{n+1}\P_A,\quad
\r_\a={\sum_{\b=1}^\a \R_\b-\a\R_{\a+1}
  \over \sqrt{\a(\a+1)}}, \;
  \p_\a= {\sum_{\b=1}^\a \P_\b-\a\P_{\a+1}
    \over \sqrt{\a(\a+1)}}.
\end{equation}
Independently of the fact that $\R, \P, \r_\a, \p_\a$ are unknown
operators, one has:
\begin{equation}  \label{Hsep}
H_{tot}=H_{CM}+H,
\quad
H_{CM}={\P^2\over{2m(n+1)}},\quad
H=\sum_{\alpha=1}^{n} \Big({ {\bf p}_\alpha^2 \over 2m}
+{m\omega^2\over{2}}{\r}_\a^2 \Big).
\end{equation}
For the equations of motion one obtains:

\noindent
$\bullet~$  Heisenberg equations
\begin{eqnarray}
&& {\dot \P}=-{i\over\hbar}[\P,H_{tot}],\quad
{\dot \R}=-{i\over\hbar}[\R,H_{tot}],\\
&&  {\dot \p}_\a=-{i\over\hbar}[\p_\a,H_{tot}],\quad
{\dot \r}_\a=-{i\over\hbar}[\r_\a,H_{tot}],\quad
\a=1,\ldots,n.\label{Heis}
\end{eqnarray}
\noindent
$\bullet~$ Hamiltonian equations
\begin{eqnarray}
&& {\dot \P}=0, \quad {\dot \R}={\P\over m(n+1)},\\
&& {\dot \p}_\a=-m\omega^2\r_\a,\quad {\dot \r}_\a={\p_\a\over m},
\quad \a=1,\ldots,n.\label{Hamil}
\end{eqnarray}
In order to find some noncanonical solutions of the $n+1$ particle quantum
system we accept the following assumptions (which also hold for any
canonical quantum system):

{\bf Assumption 1.} The CM variables commute with the internal
variables.

{\bf Assumption 2.} The CM coordinates and momenta
are canonical.

{\bf Assumption 3.}

$1^0$. The angular  momentum operators  {\bf J} ={\bf
L}+{\bf M} are generators  of the
algebra  $so(3)$  of the space rotations ({\bf L} and {\bf M}
are the angular momentum of the centre of mass and the
internal angular momentum);

$2^0$.  $H_{tot}=H_{CM}+H$  is  a  generator  of  the translations
in time $t$;

$3^0$.
      The  components  of  the total momentum
          {\bf P}  are  generators  of
          space translations;

$4^0$. {\bf K}=$m(n+1)${\bf R}~$-$~{\bf P}$t$
are generators of the acceleration.

{\bf Assumption 4.} The components of {\bf M} and $H$ are
generators of the algebra $so(3) \oplus u(1)$.  They are in the
enveloping
algebra of the internal position and momentum operators ${\bf
r}_\a ,\;
{\bf p}_\a $ and the following relations hold ($\a=1,\ldots,n, \;
j,k,l=1,2,3$):
\begin{equation}
[M_j,M_k]=i\varepsilon_{jkl}M_l, \;\;
[M_j,r_{\a k}]=i\varepsilon_{jkl}r_{\a l},\;\;
[M_j,p_{\a k}]=i\varepsilon_{jkl}p_{\a l}.
\end{equation}
Assumption 2 settles the compatibility between the CM variables.
Therefore it remains to investigate a $3n$-dimensional Wigner
quantum oscillator. Introduce in place of $\bf{r}_\a ,~\bf{p}_\a
$ new unknown operators:
\begin{equation}
a_{\alpha
k}^\pm=\sqrt{c_nm \omega \over 4 \hbar} r_{\a k} +\mu {i  \sqrt
{c_n\over 4m \omega \hbar}}p_{\a k} , \;\; k=1,2,3, \; \alpha
=1,\ldots , n, \; \mu =\pm \;or \;\mp, \; c_n\in {\bf N}.
\end{equation}
Then the internal Hamiltonian
$H$~(\ref{Hsep})  and the compatibility condition of the internal
Heisenberg equations~(\ref{Heis}) and the internal Hamiltonian
equations~(\ref{Hamil}) read:
\begin{eqnarray}
&& H={\omega \hbar\over{c_n}}\sum_{\a=1}^n \sum_{i=1}^3 \{a_{\a i}^+,
a_{\a i}^- \}, \\ &&\sum_{\b=1}^n\sum_{j=1}^3  [ \{a_{\b j}^+,a_{\b j}^-
\},a_{\a i}^\pm] =-\mu c_na_{\a i}^\pm , \quad i=1,2,3, \quad \alpha
=1,2,\ldots , n. \label{MQC} \end{eqnarray} We refer to the
condition~(\ref{MQC}) as a main quantum condition (MQC). In order to be
slightly more rigorous we introduce the following terminology.

\begin{definition}
{Definition } The (free unital) associative algebra $F(c_n)$ of
the generators $a_{\a i}^{\pm},$ $ \a =1,2, \ldots , n,$ $
i=1,2,3$ and the relations~(\ref{MQC}) is said to be (free)
oscillator algebra (FOA).
\end{definition}

The problem is to find those representations of the FOA, for which the
quantum conditions {\bf C1-C3} hold. In general this problem is an open
one. Here we list three classes of solutions, which are closely related to
three classes of Lie superalgebras.

\subsection{ Osp(1/6n) class of  solutions}

Let $F(2) $ be the (free unital) associative superalgebra with odd
generators  $a_{\a i}^{\pm}, \;  \a =1,2, \ldots , n, \;\; i=1,2,3$ and
relations ($\xi, \eta, \epsilon =\pm$) \begin{equation} \label{F1}
[\{a_{\a i}^\xi,a_{\b j}^\eta \},a_{\g k}^\epsilon]= \delta _{\a \g}\delta
_{ik} (\epsilon -\xi)a_{\b j}^\eta + \delta _{\b \g}\delta _{jk}(\epsilon
- \eta)a_{\a i}^\xi. \end{equation} The operators~(\ref{F1}) are para-Bose
operators. They satisfy Eqs.~(\ref{MQC}) with $\mu=\mp$ and $c_n=2$.
Therefore we have:

{\it Conclusion 1.}
$F(2)$ is a factor algebra of $F(c_n)$.

{\it Conclusion 2.} Any representation of  $F(2)$ is a
representation of $F(c_n)$.

{\it Remark 1.} The canonical Bose solution is from this class.

{\it Remark 2.} The solutions of Wigner belong also to this class.

{\it Proposition 1.} $F(2)$ is the universal enveloping algebra
$U[osp(1/6n)]$ of the orthosymplectic Lie superalgebra
$osp(1/6n)$. The pB operators are odd elements generating a Lie
superalgebra, which is isomorphic to $B(0/3n)\equiv
osp(1/6n)$~\cite{Ganchev}.

{\it Conclusion 3.} Any representation of $osp(1/6n)$ is a
representation of the FOA; it is a candidate for a Wigner quantum
system. The statistics of the latter is a $B$-statistics.

Unfortunately the representations  of $osp(1/6n)$ are still not
known explicitly.

\subsection{Sl(1/3n)  class of solutions~\cite{Palev7}}

 \s Let $F(3n-1)$ be the (free unital) associative
superalgebra with odd generators $a_{\a i}^{\pm}, \a =1,2, \ldots , n,
\;\; i=1,2,3$ and relations \begin{eqnarray} && [\{a_{\a i}^+,a_{\b
j}^-\},a_{\g k}^+]= -\delta_{jk}\delta_{\b \g}a_{\a i}^+
+\delta_{ij}\delta_{\a \b}a_{\g k}^+,\quad \{a_{\a i}^+,a_{\b j}^+\}=0,
\nn\\ && [\{a_{\a i}^+,a_{\b j}^-\},a_{\g k}^-]= \delta_{ik}\delta_{\a
\g}a_{\b j}^- -\delta_{ij}\delta_{\a \b}a_{\g k}^-, \qquad \{a_{\a
i}^-,a_{\b j}^-\}=0. \label{resh2} \end{eqnarray} These operators also
satisfy the MQC ($\mu =\mp, \; c_n=3n-1$) and therefore $F(3n-1)$ is a
factor algebra of the FOA $F(c_n)$.

{\it Proposition 2.} $F(3n-1)$ is (isomorphic to) the universal
enveloping algebra $U[sl(1/3n)]$ of the LS $sl(1/3n)\equiv
A(0,3n-1)$ from the class {\bf A} of the basic Lie superalgebras.

Any representation of $sl(1/3n)$ gives a representation of the
operators~(\ref{resh2}). Hence such operators do exist and the
corresponding to them statistics is an $A-$statistics.

In~\cite{Palev7} a class of representations obtained by the usual
Fock space technique were constructed. The conditions {\bf C1-C3}
lead to finite-dimensional irreducible $sl(1/3n)$ modules
$W(n,p)$, where the number $p=1,2,\ldots, $ called an order of
the statistics, characterizes the representation. Here is a list
of some properties of these statistics.

$\bullet$ The internal energy $E_k$ of the system takes
$min(3n,p)+1$ different values: $ E_k={\omega \hbar\over
{3n-1}}\Big(3np-(3n-1)k\Big), \quad k\equiv \sum_{\a =1}^n\sum_{i=1}^3
\t_{\a i}=0,1,2,\ldots,min(3n,p),\;\; \t_{\a i}=0,1. $

$\bullet$ The internal angular momentum of the composite
$(n+1)-$particle system takes all integer values between $0$ and
$n$.

$\bullet$ The internal geometry is noncommutative:
 $ [r_{\a i}, r_{\a j}]\ne 0, \quad i\ne j=1,2,3. $ The
positions of the particles cannot be localized. It turns out
however that $ [H, {\bf r}_\a^2]=0,~~~[{\bf r}_\a^2,{\bf
r}_\b^2]=0,~~~ \a, \b =1,\ldots,n. \nn $ Therefore the
oscillating "particles" move along spheres around the centre of
mass with radiuses $ |r_\a|=\sqrt{{\hbar\Big(3p-3k+\sum_{i=1}^3
\t_{\a i}}\Big) \over {(3n-1})m \omega},\;
k\equiv \sum_{\a =1}^n\sum_{i=1}^3\t_{\a i}=
0,1,2,\ldots,min(3n,p), \t_{\a i}=0,1 . $ Setting in the last
Eqs.  $k=0$, one obtains the maximal radius. Hence the diameter
of the oscillator is $ d=2\sqrt{3\hbar p\over (3n-1)m\omega}.\nn
$ The different  "particles" can stay simultaneously on spheres
with different radiuses. Their positions, however, cannot be
localized because, as mentioned, the coordinate operators do not
commute with each other.

$\bullet$ Turning to the initial $n+1$-particle system one notes
that the maximal distance between any two particles is $d$.
The system exhibits a
nuclear kind structure: all particles are locked in a small
volume $V$ (with spatial dimension $d$) around the centre of mass.
Since the coordinates do not commute, the particles are smeared
with a certain probability within $V$ and the geometry within
this volume is noncommutative.

\subsection{Sl(n/3) ($n\neq 3$) class of solutions}

Let $F(|n-3|)$ be the (free unital) associative superalgebra
with odd
generators  $a_{\a i}^{\pm}, \;  \a =1,2, \ldots , n, \;\; i=1,2,3$ and
relations ($\xi, \eta, \epsilon =\pm$) \begin{eqnarray} && [\{a_{\a
i}^+,a_{\b j}^-\},a_{\g k}^+]= \delta_{jk}\delta_{\a \b}a_{\g i}^+
-\delta_{ij}\delta_{\b \g}a_{\a k}^+,\quad~~
 \{a_{\a i}^+,a_{\b j}^+\}=0,
\nn \\
&& [\{a_{\a i}^+,a_{\b
j}^-\},a_{\g k}^-]= -\delta_{ik}\delta_{\a \b}a_{\g j}^-
+\delta_{ij}\delta_{\a \g}a_{\b k}^-, \quad
\{a_{\a i}^-,a_{\b j}^-\}=0.
\label{F3}
\end{eqnarray}
It is
straightforward to check that this operators satisfy the MQC with $\mu
=\mp$, $c_n=3-n$ for $n=1,2$ and $\mu =\pm$, $c_n=n-3$ for $n>3$.
Therefore $F(|n-3|)$ is again a factor algebra of the FOA $F(c_n)$.

{\it Proposition 3.} The operators $a_{\a i}^{\pm}$, $\a =1,\l ,n$,
$i=1,2,3$ constitute a basis in the odd subspace of the Lie superalgebra
$sl(n/3)$ and generate the whole algebra.

{\it Proof.} First we define the Lie superalgebra $sl(n/m)$.
The universal enveloping algebra
$U[gl(n/m)]$ of the general linear LS
$gl(n/m)$ is a {\bf Z}$_2-$graded associative unital superalgebra
generated by $(n+m)^2\;$ {\bf Z}$_2-$graded
indeterminates $\{e_{ij}|i,j\in [1;n+m]\}$, $deg(e_{ij})=\theta_{ij}$,
subject to the relations: \begin{equation}\label{super}
\[e_{ij},e_{kl}\]=\d_{jk}e_{il}-(-1)^{\t_{ij}\t_{kl}}\d_{il}e_{kj}, \quad
i,j,k,l\in [1;n+m]. \end{equation} The LS $gl(n/m)$ is a subalgebra of
$U[gl(n/m)]$, considered as a Lie superalgebra, with generators
$\{e_{ij}|i,j\in [1;n+m]\}$ and supercommutation relations~(\ref{super}).
The LS $sl(n/m)$ is a subalgebra of $gl(n/m)$: \begin{equation}
sl(n/m)=lin.env.\{e_{ij}, (-1)^{\t_k}e_{kk}-(-1)^{\t_l}e_{ll}| i\ne j;\;
i,j,k,l\in [1;n+m]\}. \end{equation} The odd generators of $sl(n/3)$,
namely  $e_{\a ,n+i}$, $e_{n+i, \a}$, $\a =1,\l ,n$, $i=1,2,3$ constitute
a basis in the odd part $sl_{\bar 1}(n/3)$. The even generators are all
their anticommutators: $ \{ e_{\a , n+i}, e_{n+j, \b}
\}=\d_{ij}e_{\a\b}+\d_{\a\b}e_{n+j,n+i}. $ Hence the operators  $e_{\a
,n+i}$, $e_{n+i, \a}$, $\a =1,\l ,n$, $i=1,2,3$ generate $sl(n/3)$. It is
easy to see that the odd generators satisfy the relations:
\begin{eqnarray} && [\{e_{n+i,\a },e_{\b,n+ j}\},e_{n+k,\g }]=
\delta_{jk}\delta_{\a \b}e_{n+i,\g } -\delta_{ij}\delta_{\b \g}e_{n+k,\a
},\nn \\ && [\{e_{n+i,\a },e_{\b,n+ j}\},e_{\g, n+k }]=
\delta_{ij}\delta_{\a \g}e_{\b, n+k} -\delta_{ik}\delta_{\a\b}e_{\g,
n+j},\\ && \{e_{\a ,n+i},e_{\b, n+j}\}= \{e_{n+i,\a },e_{n+j,\b }\}=0,
\;\;\a,\b,\g=1,\l,n,\;i,j,k=1,2,3. \nn \end{eqnarray} Therefore the
operators $
 a_{\a i}^+ =e_{n+i,\a }, \;\;
a_{\a i}^- =e_{\a, n+i }
$
satisfy Eqs.~(\ref{F3}), constitute a basis in the odd subspace of the Lie
superalgebra $sl(n/3)$ and generate it. Clearly these properties hold for
any other representation of $sl(n/3)$, which completes the proof.
\rule{5pt}{5pt}

{\it Conclusion 1.} The algebra $F(|n-3|)$ is (isomorphic) to the
universal enveloping algebra
$U[sl(n/3)]$ of the LS $sl(n/3)\equiv A(n-1/2)$ from the class {\bf A} of
the basic Lie superalgebras.

{\it Conclusion 2.} The operators $a_{\a i}^{\pm}$~(\ref{F3}) satisfy the
condition {\bf C2} of the definition of a Wigner quantum system.

\n
In terms of these operators
$
H={\omega \hbar\over{|n-3|}}\sum_{\a=1}^n  \sum_{i=1}^3
\{a_{\a i}^+, a_{\a i}^- \}.
$

\n
Set
$
 M_{j}=
-{i\over n}\sum_{\a =1}^n\sum_{k,l=1}^3 \varepsilon_{jkl}
\{a_{\a k}^+,a_{\a l}^- \}.
$
Then
$
[M_j,H]=0,\;
[M_j,A_k]=i\varepsilon_{jkl}A_l, \;
A_k \in \{M_i,\; r_{\a i}, \; p_{\a i} |i=1,2,3;\; \a=1,\ldots,n\}.
$
In such a way we have

{\it Conclusion 3.}
The condition {\bf C3} of the definition of a WQS
is fulfilled.

In order to satisfy also {\bf C1}
we have to define the position and the momentum
operators $\r_\a$ and $\p_\a$, corresponding to the operators~(\ref{F3}),
as linear Hermitian operators in a Hilbert space $W_{int}$. This means
that the Hermitian conjugate to $a_{\a k}^+$ should be equal to $a_{\a
k}^-$, i.e., \begin{equation}\label{conj} (a_{\a k}^+)^\dagger=a_{\a k}^-.
\end{equation} The operators~(\ref{F3}) have several representations. Here
we consider a class of irreducible modules $W(n,p)_{int}$ labeled by one
positive integer $p=1,2,\ldots$. Let \begin{equation}\label{basis}
|p;s)\equiv |p;s_1,s_2,\ldots,s_{n+2})\in W(n,p)_{int}. \end{equation}
Postulate that the set of all vectors $|p;s) $ with
\begin{equation}\label{cond} s_i\in {\bf Z}_+,\;\; i=1,2,\ldots ,n-1;
\quad s_{n}, s_{n+1}, s_{n+2}\in \{0,1\},\quad \sum_{l=1}^{n+2}s_l \leq p
\end{equation} constitute an orthonormed basis in $W(n,p)_{int}$. Let $
|p;s)_{\pm i; \pm j}$ be a vector, obtained from $|p;s)$ after a
replacement of $s_{i}$ with $s_{ i}\pm 1$ and  $s_{j}$ with $s_{j}\pm 1$.
Then the transformation of the basis under the action of the operators
$a_{\a i}^{\pm}$ read ($\a =2,\ldots ,n$, $i=1,2,3$): \begin{eqnarray} &&
a_{1i}^-|p;s)= (-1)^{s_n+\ldots  +s_{n+i-2}}s_{n-1+i}\sqrt{
p-\sum_{l=1}^{n+2} s_l +1}\;\;|p;s)_{-(n-1+i)},\\ && a_{1i}^+|p;s)=
(-1)^{s_n+\ldots +s_{n+i-2}}(1-s_{n-1+i}) \sqrt{ p-\sum_{l=1}^{n+2} s_l
}\;\;|p;s)_{+(n-1+i)},\\ && a_{\a i}^-|p;s)= (-1)^{s_n+\ldots
+s_{n+i-2}}s_{n-1+i} \sqrt{s_{\a -1}+1}\;\;|p;s)_{+(\a -1);-(n-1+i)}, \\
&&a_{\a i}^+|p;s)= (-1)^{s_n+\ldots +s_{n+i-2}}(1-s_{n-1+i}) \sqrt{s_{\a
-1}}\;\;|p;s)_{-(\a -1);+(n-1+i)}. \end{eqnarray} It is straightforward to
verify~(\ref{conj}) and hence {\bf C1} holds.

The  state space $W(n,p)_{int}$ is finite-dimensional.
The internal Hamiltonian $H$ is diagonal in the
basis~(\ref{basis})-(\ref{cond})
\begin{equation}
H|p; s) ={\omega \hbar\over
{|n-3|}}\big(3p+(n-3)(r_n+r_{n+1}+r_{n+2}) \big)
|p; s).
\end{equation}

In order to compute the eigenvalues of the internal angular
momentum one has to decompose each $sl(n/3)$ module $W(p,n)_{int}$
according to the chain
\begin{equation}
sl(n/3) \supset sl(n) \oplus sl(3) \oplus u(1).
\end{equation}
This problem will be considered elsewhere.

\bigskip
In conclusion, we have considered three different examples of WQSs
corresponding to one and the same Hamiltonian~(\ref{Ham}). The defining
conditions {\bf C1,~C2,~C3} for the systems to be WQS were satisfied with
position and momentum operators, which generate different representations
of the Lie superalgebras $osp(1/6n)$ $sl(1/3n)$ and $sl(n/3)$. Some of the
physical properties of these WQSs were shortly discussed.

\section*{Acknowledgments}

N.I.\ Stoilova is thankful to Prof.\ H.D.\ Doebner for constructive
discussions. The work was supported by the Humboldt Foundation and
the Grant N $\Phi -910$ of the Bulgarian Foundation for Scientific
Research.

\end{document}